\documentclass{iaus}
\usepackage{graphicx}
\usepackage{natbib}

\title[GWs and GRBs] 
{Gravitational waves and gamma-ray bursts}

\author[Alessandra Corsi et al.]   
{Alessandra Corsi$^1$ \\for the LIGO Scientific Collaboration and Virgo collaboration
 }
\affiliation{$^1$LIGO laboratory, California Institute of Technology\\MS 100-36, Pasadena, CA 91125 (USA) \\ email: {corsi@caltech.edu}}

\pubyear{2012}
\volume{279}  
\pagerange{xxx--yyy}
\jname{Death of Massive Stars: Supernovae and Gamma-Ray Bursts}
\editors{P. Roming, N. Kawai \& E. Pian, eds.}
\begin{document}

\maketitle
\begin{abstract}
Gamma-Ray Bursts are likely associated with a catastrophic energy release in stellar mass objects. Electromagnetic observations provide important, but indirect information on the progenitor. On the other hand, gravitational waves emitted from the central source, carry direct information on its nature. In this context, I give an overview of the multi-messenger study of gamma-ray bursts that can be carried out by using electromagnetic and gravitational wave observations. I also underline the importance of joint electromagnetic and gravitational wave searches, in the absence of a gamma-ray trigger. Finally, I discuss how multi-messenger observations may probe alternative gamma-ray burst progenitor models, such as the magnetar scenario.
\keywords{gamma rays: bursts, gravitational waves, stars: neutron, supernovae: general}
\end{abstract}
\firstsection
\section{Introduction}
During the last 15 years, thanks to satellite missions like \textit{BeppoSAX} \citep{Boella1997}, \textit{Swift} \citep{Gehrels2004}, and \textit{Fermi} \citep{Atwood2009,Meegan2009}, our progress in the understanding of gamma-ray bursts (GRBs) has been quite spectacular. We now know that GRBs are cosmological events related to a catastrophic energy release in stellar mass objects. Energy dissipation within a highly relativistic ``fireball'', presumably emitted in the form of a jet \citep[e.g.,][]{Rhoads1999,Sari1999,Kumar2000,Frail2001}, is believed to power the observed $\gamma$-ray flash (prompt emission) and the subsequent ``afterglow'' \citep[e.g.,][]{Blandford1976,Rees1992,Meszaros1993a,Meszaros1993b,Piran2004,Meszaros2006}. 

Traditionally, GRBs have been divided in two main categories, long and short ones \citep[e.g.][]{Kou1993}, depending on the duration of their prompt $\gamma$-ray emission ($\lesssim 2$\,s for the short bursts, $\gtrsim 2$\,s for the long ones). These two populations of bursts are thought to be related to two different progenitor models: ``collapsars'' for the long-soft bursts \citep[e.g.,][ and references therein]{Woosley1993,Mac1999,Piran2004,Meszaros2006,Woosley2006}, and the merger of binary systems of compact objects such as neutron star (NS)-NS or black-hole (BH)-NS systems, for the short-hard ones \citep[e.g.,][]{Eichler1989,Narayan1992,Janka1999,Bel2002,Rosswog2005,Bel2006,Faber2006}. 

Despite the recent issues raised in the classification of short and long GRBs based solely on their prompt emission properties \citep[e.g,][]{Zhang2009}, the general picture of two classes of bursts related to two main progenitor models, still holds. The collapsar scenario is observationally supported by the fact that long GRBs occur in galaxies with high specific star formation rate \citep[e.g.,][]{Chris2004,Castro2006,Fruchter2006,Levesque2010}, and that at least some long GRBs have been observed to be associated with core-collapse supernovae (SNe) of rare type \citep{Galama1998,Berger2003a,Hjorth2003,Malesani2004,Campana2006,Pian2006}. Indeed, the question of what makes some stars die as SNe and some other as relativistic GRBs, is not solved yet \citep[e.g.,][and references therein]{Woosley2006}. Indications that progenitors of short bursts belong, on average, to old stellar populations with a typical lifetime of several Gyr \citep{Barthelmy2005,Berger2005,Ghe2005,Villa2005,Bloom2006,Gal-Yam2008,Richard2008}, provide support to the binary merger scenario.

Collapsars and binary mergers leading to the formation of a BH plus an accretion disk \citep[e.g.,][]{Woosley1993,Fryer1996,Mac1999,Rosswog1999,Ruffert1999,Narayan2001}, have the potential to power the GRB fireball via the energy released from the accretion of the disk onto the newly formed BH. While the formation of a BH plus disk system is common to both progenitor models, the more compact scale of the NS-NS (or BH-NS systems), and the less massive debris left over after merger, are invoked to explain the shorter duration and the smaller isotropic energies of these bursts with respect to long ones.

Within the standard fireball model, once the fireball is launched from the central engine, the observed radiation is explained as synchrotron and/or inverse Compton emission from electrons accelerated in internal and external shocks \citep[e.g.,][]{Sari1997,Kobayashi1997,Sari1998,Granot1999,Dermer2000,Sari2001}, taking place at distances $\gtrsim 10^{13}$\,cm from the central source. High-energy (GeV) tails observed in some GRBs have challenged the internal-external shock fireball model in its simplest formulation \citep{Hurley1994,Baring1997,Abdo2009,Kumar2009,Ackermann2010,Corsi2010b,DePasquale2010,Ghirlanda2010,Giuliani2010,Abdo2011,Asano2011,Meszaros2011,Toma2011,Zhang2011}. However, it remains true that the electromagnetic emission from GRBs, being produced at large distances from the central engine, provides indirect information on the progenitor. On the other hand, gravitational waves (GWs) emitted from the progenitor could directly probe its nature.
\section{GRB-triggered searches for GWs}
\label{triggered}
Being related to catastrophic events involving stellar-mass objects, GRBs are good candidates for the detection of GWs \citep{Kochanek1993,Finn1999,vanPutten2001,vanPutten2002,Kobayashi2003a}. Coalescing binaries, thought to be associated with short bursts, are one of the most promising GW sources \citep[e.g.,][and references therein]{Phinney1991,Cutler1993,Zhuge1994,Flanagan1998,Rates2010,Shibata2011} for detectors like the Laser Interferometer Gravitational-Wave Observatory \citep[LIGO; ][]{Abbott2009} and Virgo \citep{Acernese2008a,Accadia2011}. For such systems, a chirp signal should be emitted in GWs during the in-spiral, followed by a burst-type signal associated with the merger, and subsequently a signal from the ring-down phase of the newly formed BH \citep[e.g.,][and references therein]{Echeverria1989,Kobayashi2003a,Berti2009}. The last, initially deformed, is expected to radiate GWs until reaching a Kerr geometry \citep{Kobayashi2003a}. 

In the collapsar scenario, relevant for long GRBs, the high rotation required to form the centrifugally supported disk that powers the GRB, should produce GWs via bar \citep[e.g.,][]{Houser1994,New2000,Baiotti2007,Dimmel2008} or fragmentation instabilities that might develop in the collapsing core \citep[see e.g.,][for recent reviews]{Fryer2003,Ott2009} and/or in the disk \citep{Davies2002,Fryer2002,Kobayashi2003a,Piro2007}. Moreover, asymmetrically infalling matter is expected to perturb the final BH geometry, leading to a ring-down phase \citep{Echeverria1989,Kobayashi2003a}. 

LIGO and Virgo have been carrying out electromagnetically triggered searches for GWs \citep{Abbott2005,Abbott2007,Acernese2007,Abbott2008a,Abbott2008b,Acernese2008,Abbott2009a,Abbott2010,Abbott2010a,Abadie2011,Abbott2012} over the past decade \citep[for bar detectors electromagnetically triggered searches, see e.g. ][]{Astone1999,Astone2002,Astone2005,Baggio2005}. The LIGO Scientific Collaboration operates two LIGO observatories in the U.S. along with the GEO600 detector \citep{Grote2010} in Germany. Together with Virgo, located in Italy, they form a detector network capable of detecting GW signals arriving from all directions. 

GRB-triggered searches for GWs by LIGO and Virgo have targeted both the chirp signal expected in the case of short GRBs during the NS-NS or BH-NS in-spiral, and short unmodeled pulses of GWs that may be expected during the merger/collapse, and ring-down phases of short/long GRBs \citep{Abbott2005,Acernese2007,Abbott2008a,Abbott2008b,Acernese2008,Abbott2010,Abbott2010a,Abbott2012}. These searches have adopted on-source time windows of few minutes (long GRBs) or few seconds (short GRBs) around the GRB trigger time. In fact, for long GRBs, the time delay between the GW signal and $\gamma$-ray trigger is thought to be dominated by the time necessary for the fireball to push through the stellar envelope of the progenitor \citep[10-100\,s;][]{Zhang2004}. On the other hand, for short GRBs, the NS-NS/BH-NS merger is believed to occur quickly, and be over within a few seconds (naturally accounting for the short nature of these bursts). It is estimated that triggered searches for GWs in few minutes time-windows yield a factor of $\approx 2$ improvement in sensitivity with respect to untriggered ones \citep{Kochanek1993}. 

The most exciting results from LIGO GRB-triggered searches of GWs are probably represented by the cases of the short GRBs 070201 \citep{Abbott2008b,Ofek2008} and 051103 \citep{Abbott2012,Hurley2010}, whose error boxes overlap with nearby galaxies (M31 for GRB\,070201; M81 for GRB\,051103). A NS-NS binary merger scenario occurring in such hosts was excluded by LIGO with rather high confidence \citep{Abbott2008b,Abbott2012}. However, the possibility that GRB\,070201 and GRB\,051103 are related to (extra-galactic) soft gamma-ray repeaters (SGR) giant flares \citep[for a recent review, see][and references therein]{Mereghetti2008}, could not be ruled out. Indeed, LIGO upper-limits for short unmodeled pulses of GWs from GRB\,070201 and GRB\,051103, are above the maximum GW energy emissible in SGR giant flares \citep{Ioka2001,Corsi2011}. 
\section{GW-triggered searches for GRBs}
A very appealing prospect is represented by the possibility of using GWs to trigger electromagnetic (radio, optical, X-ray) follow-ups of GW sources \citep[e.g.,][]{Sylvestre2003,Bloom2009,Metzger2012}.  The discovery of off-axis optical or radio afterglows \citep{Meszaros1998,Granot2002,Janka2006,van2011} triggered via the (non-beamed) GW emission from the GRB progenitors, would yield a dramatic confirmation of the ``jet model'', map out the beaming distribution, and provide fundamental inputs to models of relativistic outflows. Radio follow-ups, in particular, are an effective tool to identify relativistic and mildly relativistic outflows \citep[e.g.,][]{Kulkarni1998,Soderberg2010,Nakar2011} in the absence of a $\gamma$-ray trigger.

Finding electromagnetic counterparts to GW triggers is technically challenging due to imperfect localization of the GW signal and uncertainty regarding the relative timing of the GW and electromagnetic emissions. The localization of LIGO-Virgo triple-coincidence GW triggers can yield error-areas of $\sim 100$\,deg$^{2}$, possibly spread over different patches of the sky \citep[see e.g. Fig. 3 in][]{Abbott2012}. The problem of following-up with optical (or radio, or X-ray) telescopes such a large error-area can be partially mitigated by: (i) restricting the search for electromagnetic counterparts to transients in nearby galaxies (within the LIGO-Virgo horizon distance); (ii) by noticing that the most promising electromagnetic counterparts of GW events detectable by LIGO and Virgo are expected to be ``exotic'' (rare) ones \citep[e.g., the orphan afterglow of a GRB, and/or the ``kilonova'' from a binary merger - see][]{Kulkarni2005,Metzger2010}.

In 2009-2010, LIGO and Virgo, together with partner electromagnetic observatories, performed their first ``LOOC-UP'' - Locating and Observing Optical Counterparts to Unmodeled Pulses of gravitational waves - experiment \citep[][and references therein]{Kanner2008,Abbott2012}. At the time, there were two operating LIGO interferometers \citep{Abbott2009}, each with 4-km arms (one near Hanford, Washington, the other in Livingston Parish, Louisiana). The Virgo 3-km arms detector \citep{Acernese2008a,Accadia2011} located near Cascina (Italy), was also operating. The LOOC-UP search has established a baseline for low-latency analyses with the next-generation GW detectors \citep[Advanced LIGO and Advanced Virgo;][]{Acernese2009,Harry2010}. The collaboration between GW and electromagnetic observatories is likely to continue to develop over the next few years, as the scientific community gets ready for a global network of advanced GW detectors.
\section{GRBs and magnetars: prospects for multi-messenger studies}
The forthcoming years may see the development of new GW searches in coincidence with GRBs, aimed at answering some of the questions opened by recent observations. In particular, a compelling result from \textit{Swift} has been the discovery that the ``normal'' power-law behavior of long GRB X-ray light curves is often preceded at early times by  a steep decay, followed by a shallower-than-normal decay \citep[e.g. ][]{Nousek2006,Zhang2006}. The steep-to-shallow and shallow-to-normal decay transitions are separated by two break times,  $100~{\rm s}\lesssim T_{break,1}\lesssim 500$~s and $ 10^{3}$~s~$\lesssim T_{break,2}\lesssim 10^{4}$~s. It has been suggested that the shallow phase may be attributed to a continuous energy injection by a long-lived central engine, with progressively reduced activity \citep{Zhang2006}.

Newborn magnetars, besides being candidate GRB progenitors \citep[e.g.,][]{Usov1992,Thompson1994,Bucciantini2007,Metzger2007}, have also been proposed to account for shallow decays or plateaus observed in GRB light curves \citep[][]{Dai1998,Zhang2001,Fan2006,Yu2007,Metzger2008,Xu2009,Row2010}. Independent support for the magnetar scenario comes from the observation of SN\,2006aj,  associated with the nearby sub-energetic GRB\,060218, suggesting that the SN-GRB connection may extend to a much broader range of stellar masses than previously thought, possibly involving two different mechanisms: a ``collapsar'' for the more massive stars collapsing to a BH, and a newborn (highly-magnetized) NS for the less massive ones \citep{Mazzali2006}.

Several studies have shown how magnetars dipole losses may indeed explain the flattening observed in GRB afterglows \citep[][]{Dai1998,Zhang2001,Fan2006,Yu2007,DallOsso2011,Bernardini2012}. \citet{Corsi2009} have explored a scenario in which the newly born magnetar left over after the GRB explosion undergoes a secular bar-mode instability \citep{LaiShapiro95}, thus producing a bar-like GW signal associated to the electromagnetic plateau, potentially detectable by the advanced ground-based interferometers like LIGO and Virgo (up to distances of $\sim 100$\,Mpc). Compared to current analyses that GW detectors are carrying out (see Section \ref{triggered}), this scenario \citep{Corsi2009} involves a new class of GW signals, with a longer duration ($10^3-10^4$ s) and a different frequency evolution. Data analysis techniques for the search of longer duration GW signals possibly applicable to GRB searches, are being developed \citep[e.g.,][]{Thrane2011}.
\section{Prospects and conclusions}
The LIGO interferometers are being upgraded to the next-generation Advanced detectors \citep{Harry2010}, that are expected to become operational around 2015. Virgo will also be upgraded to become Advanced Virgo \citep{Acernese2009}. Additionally, the new LCGT detector \citep{Kuroda2010} is being built in Japan. These advanced detectors are expected to detect compact binary coalescences, possibly at a rate of dozens per year after reaching design sensitivity \citep{Rates2010}, so that the short-GRB progenitor scenario may finally be probed directly. Long-standing open questions (e.g., is the jet model for GRBs correct? Why do some massive stars die as SNe and others as relativistic GRBs?), or other issues raised by more recent observations (such as the difficulties in the long-short GRB classification; the role of magnetars as GRB progenitors; the link between long GRBs and SGRs, etc.), will greatly benefit from joint GW studies. The advanced GW detectors will provide a totally new view of the transient sky \citep{Marka2010,Marka2011}. The prospects for this new era of astronomy are exciting, and promise a return of big scientific impact.
\vspace{0.3cm}

\footnotesize{\textbf{Acknowledgments}\\
LIGO was constructed by the California Institute of Technology and Massachusetts Institute of Technology with funding from the National Science Foundation and operates under cooperative agreement PHY-0757058. This paper has LIGO Document Number LIGO-P1200042.}

\begin{discussion}
\discuss{ASTRAATMADJA}{Is the angular resolution of the GW detectors good enough to search for an electromagnetic counterpart? Do you also intend to look for neutrino signals?}
\discuss{CORSI}{The error-area for triple coincidence GW events from the LIGO-Virgo network is $\sim100$ deg$^2$, much bigger than e.g. the $\approx 2''$ FWHM of a telescope like the Palomar 48-inch \citep{Rau2009}. While a large number of optical transients is expected to be found in the GW error-area, the problem can be mitigated by selecting only the most promising for a GW detection (in nearby galaxies and likely of  ``exotic'', rare type). Joint searches for GWs and high energy neutrinos (though, currently, not specifically within the LOOC-UP experiment) are indeed being performed \citep[see e.g.,][and references therein]{Bou2012}. }
\discuss{METZGER}{In the magnetar scenario proposed for explaining GRB plateaus, can sufficiently rapid rotation be maintained in the presence of enhanced early spin-down by neutrino emission?}
\discuss{CORSI}{Sufficiently high rotation should be maintained to explain the observed plateaus: typically, a $(1-5)$\,ms magnetar with $B\sim (1-10)\times10^{14}$\,G is required from modeling of the X-ray light curves with plateaus \citep[e.g.,][]{Zhang2001,Yu2007,Xu2009,DallOsso2011}. As you have shown \citep{Metzger2007}, enhanced spin-down by neutrino emission at earlier timescales may be an issue, but likely only for the shortest periods and highest magnetic fields in these ranges.} 
\end{discussion}

\end{document}